\documentclass[]{article}
\usepackage{multicol}
\usepackage{apj}
\def\lesssim{\mathrel{\hbox{\rlap{\hbox{\lower4pt\hbox{$\sim$}}}\hbox{$<$}}}}
\def\gtrsim{\mathrel{\hbox{\rlap{\hbox{\lower4pt\hbox{$\sim$}}}\hbox{$>$}}}}
\newcommand{\dnsig}{D_n-\sigma}
\newcommand{\mincir}{\raise -2.truept\hbox{\rlap{\hbox{$\sim$}}\raise5.truept
\hbox{$<$}\ }}
\newcommand{\magcir}{\raise -2.truept\hbox{\rlap{\hbox{$\sim$}}\raise5.truept
\hbox{$>$}\ }}
\newcommand{\siml}{\raise -2.truept\hbox{\rlap{\hbox{$\sim$}}\raise5.truept
\hbox{$<$}\ }}
\newcommand{\simg}{\raise -2.truept\hbox{\rlap{\hbox{$\sim$}}\raise5.truept
\hbox{$>$}\ }}
\newcommand{\be}{\begin{equation}}
\newcommand{\ee}{\end{equation}}
\newcommand{\ba}{\begin{eqnarray}}
\newcommand{\ea}{\end{eqnarray}}
\newcommand{\lb}{{\left<\right.}}
\newcommand{\rb}{{\left.\right>}}
\newcommand{\br}{\mbox{\bf r}}
\newcommand{\hm}{\,h^{-1}{\rm Mpc}}
\newcommand{\vel}{\,{\rm km\,s^{-1}}}

\newcommand {\h} {$h^{-1}$ Mpc $ \;$}

\begin{document}

\vspace{15mm}                                                       
\begin{center}
\uppercase{ENEAR Redshift-Distance Survey: Cosmological Constraints}\\
\vspace*{1.5ex} 
{\sc
S. Borgani$^{1,2}$, M. Bernardi$^{3,4}$, L. N. da Costa$^{4,5}$
G. Wegner$^6$, M. V. Alonso$^7$,\\
C. N. A. Willmer$^8$, P. S. Pellegrini$^9$ \& M. A. G. Maia$^9$
}\\
\vspace*{1.ex}
{\small
$^{1}$ INFN, Sezione di Trieste, c/o Dipartimento di
Astronomia, Universit\`a di Trieste, via Tiepolo 11,
I-34131 Trieste (Italy)\\
$^{2}$ INFN, Sezione di Perugia, c/o Dipartimento di
Fisica dell'Universit\`a, via A. Pascoli, I-06123 Perugia (Italy)\\
$^{3}$ The University of Chicago, 5640 South Ellis
Avenue, Chicago, IL 60637 (U.S.A.)\\
$^{4}$ European Southern Observatory, Karl
Schwarzschild Str. 2, D--85748 Garching b. M\"unchen (Germany)\\
$^{5}$ Observat\'orio Nacional, Rua Gen. J. Cristino
77, S\~ao Cristov\~ao, Rio de Janeiro (Brazil)\\
$^{6}$ Dept. of Physics and Astronomy, Dartmouth College,
Hanover, NH 03755 (U.S.A.)\\
$^{7}$ Observatorio Astr\'onomico de
C\'ordoba,  Laprida  854, C\'ordoba, 5000 (Argentina)\\
$^{8}$ UCO/Lick Observatory, University of California,
1156 High Street, Santa Cruz, CA 95064 (U.S.A.)\\
$^{9}$ Observat\'orio Nacional, Rua General Jos\'e Cristino
77, Rio de Janeiro, R. J., 20921 (Brasil)}
\end{center}

\vspace*{-6pt}

\begin{abstract}
We present an analysis of the ENEAR
sample of peculiar velocities of field and cluster elliptical
galaxies, obtained with $D_n$--$\sigma$ distances. We use the velocity
correlation function, $\psi_1(r)$, to analyze the statistics of the
field--object's velocities, while the analysis of the cluster data is
based on the estimate of their r.m.s. peculiar velocity, $V_{\rm
rms}$. The results are compared with predictions from cosmological
models using linear theory. The statistics of the model velocity field
is parameterized by the amplitude, $\eta_8=\sigma_8 \Omega_m^{0.6}$,
and by the shape parameter, $\Gamma$, of the CDM--like power
spectrum. This analysis is performed in redshift space, so as to
circumvent the need to address corrections due to inhomogeneous
Malmquist bias and to the redshift cutoff adopted in the sample
selection.  From the velocity correlation statistics we obtain
$\eta_8=0.51^{+0.24}_{-0.09}$ for $\Gamma=0.25$ at the $2\sigma$ level
for one interesting fitting parameter. This result agrees with that
obtained from a similar analysis of the SFI I-band Tully--Fisher (TF)
survey of field Sc galaxies. Even though less constraining, a
consistent result is obtained by comparing the measured $V_{\rm rms}$
of clusters to linear theory predictions. For $\Gamma=0.25$ we find
$\eta_8=0.63_{-0.19}^{+0.22}$ at $1\sigma$. Again, this result agrees,
within the uncertainties, with that obtained from the SCI cluster
sample based on TF distances. Overall, our results point toward a
statistical concordance of the cosmic flows traced by spirals and
early-type galaxies, with galaxy distances estimated using TF and
$D_n$--$\sigma$ distance indicators, respectively.

%
\vspace*{6pt}
\noindent
{\em Subject headings: }
Cosmology: observations -- cosmology: theory
-- galaxies: distances and redshifts -- large-scale structure of
universe.
\end{abstract}

\begin{multicols}{2} 

\section{INTRODUCTION}
\label{intro}
The analysis of the peculiar velocities of galaxies and clusters is
one of the most promising ways to investigate the amplitude of cosmic
density perturbations on $\sim$ 100\h1Mpc scales (e.g., Strauss \&
Willick 1995). The importance of cosmic flows for cosmology has
motivated a two-decade long effort of building large and homogeneous
redshift-distance samples of galaxies and clusters. Analyses of early
redshift-distance surveys of spirals (Aaronson et al. 1982) and of
early-types (e.g. Lynden-Bell et al. 1988), even though leading to the
development of several statistical methods of analyzing peculiar
velocity data, left many issues unresolved, primarily because they
were based on relatively small and shallow data sets.  Recently, a
second-generation of redshift-distance surveys has become available
involving high-quality data and significantly larger samples of both
spirals (Mathewson, Ford \& Buchhorn 1992; Haynes et al. 1999a,b) and
early-types (da Costa et al. 2000a).  The existence of these new
samples has raised the hope that some of the discrepancies found in
earlier analyses may soon be settled.  Indeed, the analyses of the
different all-sky catalogs of peculiar velocity data currently
available such as Mark~III (Willick et al. 1997) and SFI (e.g. da Costa
et al. 1996; Giovanelli et al. 1998), lead to a roughly consistent
picture of the peculiar velocity field and the local mass distribution
(Dekel et al. 1999). However, some quantitative disagreements still
remain ranging from the amplitude of the bulk velocity (da Costa et al.
1996; Giovanelli et al. 1998; Dekel et al. 1999), to estimates of the
parameter $\beta=\Omega_m^{0.6}/b$ (e.g., Davis, Nusser \& Willick
1996; Zaroubi et al. 1997; da Costa et al. 1998; Willick \& Strauss
1998; Freudling et al. 1999; Borgani et al. 2000), where $\Omega_m$ is
the cosmological matter density parameter and $b$ is the linear galaxy
biasing factor.  It is important to emphasize that the two most
important catalogs currently in use, Mark~III and SFI, consist of
combinations of distinct data sets covering different parts of the sky
and therefore could be susceptible to subtle systematic effects.  Both
catalogs also rely predominantly on Tully-Fisher distances of spiral
galaxies and we should note that earlier statistical comparisons of
the velocity fields derived from $\dnsig$ and TF distances found
significant differences between them (e.g., G\'orski et al. 1989;
Tormen et al. 1993). There have also been claims of significant
differences, larger than expected from the estimated errors, between
cluster distances estimated using galaxies of different morphological
types (e.g. Mould et al. 1991; Scodeggio, Giovanelli \& Haynes 1998).

In this context, the recently completed all-sky redshift--distance
survey of early-type galaxies (da Costa et al. 2000a, hereafter ENEAR),
probing a volume comparable to that of the existing catalogs of
peculiar velocity data, is a welcome addition.  The ENEAR galaxies
sample different regions of space and density regimes; the peculiar
velocities are measured using an independent distance indicator; and
the distances are based on separate types of observations, reduction
techniques and corrections. Finally, the ENEAR sample has well defined
selection criteria, the completeness of the observations is uniform
across the sky and the data, mostly new measurements by the same
group, are in a homogeneous system.

The present Letter has the twofold aim of comparing global statistical
quantities, which describe the velocity fields traced by the TF and
$\dnsig$ distance indicators, and of placing constraints on the nature
of the fluctuation power--spectrum. Our analysis is based on the
velocity correlation statistics and the r.m.s. one-dimensional
peculiar velocity of clusters. These statistics were used by Borgani
et al. (2000, B00 hereafter) and Borgani et al. (1997, B97 hereafter)
to analyze the SFI sample of field spirals and the SCI sample of
cluster spirals (Giovanelli et al. 1997), respectively. In this paper,
the same analysis is carried out for the ENEAR sample of field
galaxies and groups and for ENEAR clusters (hereafter ENEARc; Bernardi
et al. 2000, in preparation).

\section{THE DATA}
\label{sample}
The ENEAR sample contains 1359 ellipticals brighter than $m_B=14.5$
with $\dnsig$ measured distances and 569 cluster galaxies in 28 clusters
(ENEARc). Galaxies have been objectively assigned to groups and
clusters using the information available from complete redshift
surveys sampling the same volume.  Our analysis is performed in
redshift--space so as to avoid correcting for inhomogeneous Malmquist
bias and the redshift cutoff adopted in the sample
selection. Therefore, we use the inverse $\dnsig$ template derived by
Bernardi et al. (2000, in preparation) combining all the cluster
data. Below, we limit our analysis to objects within $cz=6000\vel$, so
as to exclude those with very uncertain velocity measurements. This
sub-sample consists of 355 field galaxies and 223 groups.  In the
cluster sample analysis we only consider the 20 clusters with
$cz\leq6000\vel$.  Of these, we discard the clusters CEN~30 and
CEN~45; these systems lie along the same line-of-sight and are close
in redshift-space making the assignment of galaxies to individual
systems difficult. They are also suspected to form a bound system
(Lucey \& Carter 1988) and their large peculiar velocities observed
may be due to non--linear effects. We also pay special attention to two
other groups, AS714 and AS753, both with large peculiar velocities
($\sim$ 900~$\vel$). These systems lie in the region of the Great
Attractor and may also be subject to non--linear dynamical
interactions.  Below we discuss the impact of including or excluding
these two systems in the analysis.

\section{THE VELOCITY CORRELATION STATISTICS}
\label{correlation}
Our analysis of the velocity correlation statistics follows closely that
presented in B00. 
We refer to that paper for a more
thorough discussion.  We use the velocity correlation estimator
originally introduced by G\'orski et al. (1989, G89 hereafter): 
\be
\psi_1(r)\,=\,{\sum_{|\br_i-\br_j|=r} u_iu_j\cos\vartheta_{ij}
\over
\sum_{|\br_1-\br_j|=r} \,\cos^2 \vartheta_{ij}}\,,
\label{eq:psi1} 
\ee  
where $\vartheta_{ij}$ is the angle between the direction of the
$i$-th and the $j$-th object and the sums are over all the 
pairs at separation $r$ in redshift space. In eq.(\ref{eq:psi1}) $u_i$
is the radial peculiar velocity of the $i$--th object and we assign
equal weight to all objects, so as to minimize the effect of cosmic
variance (see the discussion in B00).
The average of $\psi_1(r)$ over an ensemble of cosmic flow realizations
is $\Psi_1(r)=\lb \psi_1(r)\rb\,=\,{\cal
A}(r)\Psi_\parallel(r)+\left[1-{\cal A}(r)\right]\Psi_\perp(r)$,
where $\Psi_\parallel$ and $\Psi_\perp$ are the radial and transverse
correlation functions of the three--dimensional peculiar velocity field
(see G89). In linear theory, they are connected to the power--spectrum
of density fluctuations, $P(k)$, according to  
\ba 
\Psi_\parallel(r) & =
& {f(\Omega_m)^2\,H_0^2\over 2\pi^2}\,\int dk \,P(k)\,\left[j_0(kr) -
2{j_1(kr)\over kr}\right ]\,; \nonumber \\ 
\Psi_\perp(r) & = &
{f(\Omega_m)^2\,H_0^2\over 2\pi^2}\,\int dk \,P(k)\,{j_1(kr)\over kr}\,,
\label{eq:psi} 
\ea where $j_i(x)$ is the $i$-th order spherical Bessel function and
$f(\Omega_m) \simeq \Omega_m^{0.6}$. The quantity ${\cal A}(r)$ is a
moment of the selection function of the sample, which is fully
specified by the spatial distribution of the objects in the sample
(e.g., G\'orski et al. 1989; B00). Therefore, the model
$\psi_1$ can be computed taking into account the specific sampling
through the ${\cal A}(r)$ function.  The velocity correlation function
$\psi_1(r)$ for the ENEAR sample is plotted in Figure~1 up to
$r=3500\vel$, for all objects within $cz=6000\vel$. This separation
range has been shown by B00 to be that where $\psi_1$ is more stable
for the SFI sample, which has about the same size as ENEAR. We choose
the bin size to be 500$\vel$ in order to keep these errors relatively
small within each separation bin. We verified that final constraints
on the model parameter are left unchanged by halving the bin
width. For the purpose of comparing ENEAR and SFI results, we show in 
Figure 1 only the statistical errors due to the internal noise of
the data set, which have been estimated as follows. At
the position of each galaxy we add to the observed peculiar velocity a
random component which is drawn from a Gaussian distribution having
r.m.s. dispersion equal to the observational error reported for that
object. Velocity correlations are then computed for 1000 realizations
of this perturbed data set and errors on $\psi_1$ are estimated at
each separation from the scatter among these realizations. Cosmic
variance must not be included here, since ENEAR and SFI probe cosmic
flows within the same region of the Universe.
\newpage
\includegraphics{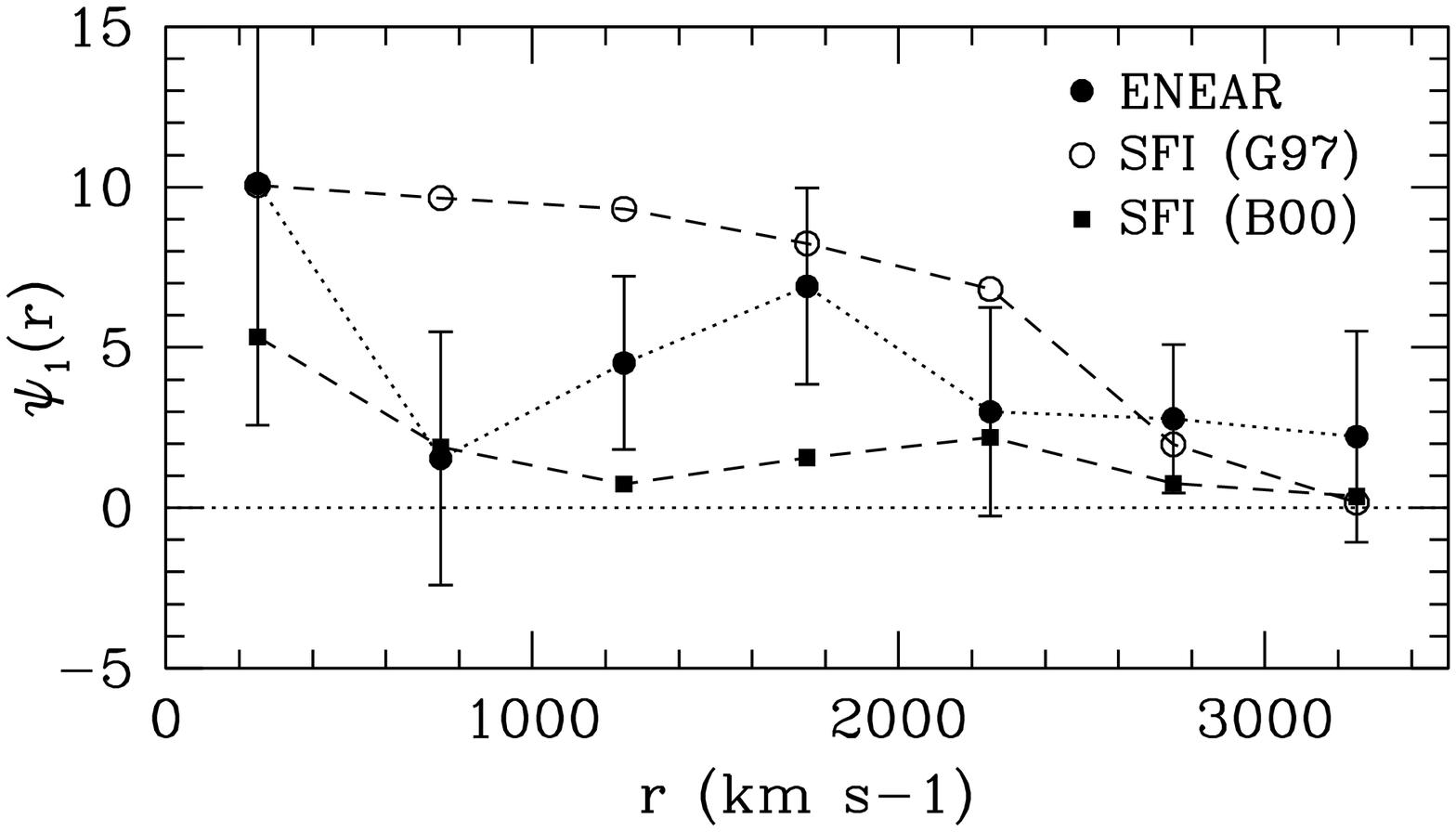}
$\ \ \ \ \ \ $\\
\vspace{5.1truecm}
$\ \ \ $\\
{\small\parindent=3.5mm {Fig.}~1.--- The velocity correlation
function, $\psi_1(r)$ (in units of $10^4\vel$), for the ENEAR sample
(filled circles). Open circles and squares are the results for the SFI
sample, as derived by B00 from two different zero--point calibrations
of the TF relation. Errorbars, that for reasons of clarity are only
reported for ENEAR, are the 1$\sigma$ uncertainties from the internal
sample noise (see text).}

\vspace{5mm}

Remarkably the $\psi_1$ velocity correlation of the ENEAR sample falls
just between the two SFI estimates, based on the two zero--point
calibrations of the TF relation presented by B00. This result
contrasts with the disagreement originally found by G89 between
spirals (Aaronson et al. 1982) and ellipticals (Lynden-Bell et
al. 1988).  We use the ENEAR velocity correlation function to place
constraints on cosmological models by following the same procedure
discussed in B00, and only briefly summarized here. We run N--body
simulations for different cosmological models, and extract from each
of them a fairly large number (256) of independent mock ENEAR
samples. In each mock sample, ``galaxies'' are placed in the same
positions as in the real sample. The peculiar velocity of each mock
galaxy is then perturbed with a Gaussian--distributed component
associated to the observational error of its real counterpart. With
this procedure, each set of mock samples includes both cosmic variance
and statistical noise. Therefore, we estimate the elements of the
covariance matrix $ {\cal
C}^{ij}=N_{mock}^{-1}\sum_{l=1}^{N_{mock}}\left(\psi_{1,l}^i-
\bar\psi_1^i\right)\, \left(\psi_{1,l}^j-\bar\psi_1^j\right)$, where
$\psi_{1,l}^i$ is the value of the velocity correlation function at
the $i$--th separation bin for the $l$--th mock sample, and
$\bar\psi_1^i$ is its ensemble average. Based on this approach, B00
showed that (a) linear theory provides a good description of the
velocity correlation statistics of the N--body simulated samples; (b)
the relative amount of covariance, i.e. the values of
$C^{ij}/\psi_1^i\psi_1^j$, is independent of the underlying
cosmology. 

Based on these results, we compute a grid of linear--theory
model predictions for $\psi_1(r)$, as well as the elements of the
corresponding covariance matrix expected for a sample the size of
ENEAR. We assume the power spectrum expression $P(k)=A\,k\,T^2(k)$,
where the transfer function, $T(k)$, is assumed to have the CDM--like
form with the $k$--dependence specified by the shape parameter
$\Gamma$. The amplitude of $P(k)$ is expressed in terms of $\sigma_8$,
the r.m.s. fluctuation amplitude within a sphere of 8$\hm$. Therefore,
following eq.(\ref{eq:psi}), $\psi_1(r)$ is entirely specified by the
two parameters $\Gamma$ and $\eta_8=\sigma_8\Omega_m^{0.6}$.  In order
to derive constraints on these parameters, we compute the weighted
$\chi^2$ between the ENEAR correlation function, $\psi^{ENEAR}_1$, and
that from model predictions, $\psi^{mod}_1$, taking into account the
covariance terms.  The probability for model rejection is estimated,
from the value of $\Delta\chi^2=\chi^2-\chi^2_{min}$, assuming a
$\chi^2$ statistic, where $\chi^2_{min}$ is the absolute minimum
value.

In Figure 2 we plot the iso--$\Delta\chi^2$ contours corresponding to
$1\sigma$ and $2\sigma$ confidence levels. The degeneracy of the
constraint in the $\eta_8$--$\Gamma$ plane is due to the fact that the
coherence of the flow on a given scale depends not only on the overall
amplitude of the power-spectrum but also on its slope. This is because
peculiar velocities are generated non--locally, so that coherence of
the flow on a given scale can be associated either to fluctuations on
comparable (large $\eta_8$ and $\Gamma$) or on much larger scales
(small $\eta_8$ and $\Gamma$). 
Fixing $\Gamma=0.25$, consistent with galaxy clustering
data (e.g., Dodelson \& Gaztanaga 1999), we find
$\eta_8=0.51^{+0.24}_{-0.09}$ at the $2\sigma$ level for one
interesting fitting parameter. We verified from the analysis of the
mock samples that redshift--space distortions have a negligible effect
in the estimate of $\psi_1(r)$, with the resulting constraints on
$\eta_8$ being affected at most by about 5\%.  As expected from the
comparison shown in Fig. 1, this result is in good agreement with that
derived by B00 from the analysis of the SFI TF-survey of spiral
galaxies. Therefore, we confirm that, for reasonable values of the
power spectrum shape, the velocity correlation statistics favor small
power--spectrum amplitudes. Although at variance with other analysis
of velocity fields based on maximum likelihood analysis of the
velocity correlation statistics (e.g. Zaroubi et al. 1998; Freudling
et al. 1999; see the discussion in B00), this result agrees with
independent constraints on the amplitude of the power spectrum, like
those imposed by the number density of local galaxy clusters (e.g.,
Eke et al. 1996; Girardi et al. 1998).

\section{The r.m.s. VELOCITY OF CLUSTERS}
\label{clusters}
The r.m.s. peculiar velocity of clusters has been used by several
authors as further means to set constraints on cosmological parameters
(e.g., Moscardini et al. 1996; Bahcall \& Oh 1996; B97; Watkins
1997). This analysis is repeated here for the ENEAR cluster sample.
The observational estimate from the ENEAR sample ranges from $V_{\rm
rms}^{\rm obs}=450\pm 73 \vel$ to $V_{\rm rms}^{\rm obs}=470\pm 68
\vel$ for the samples of 16 and 18 clusters, respectively, defined in
Section~\ref{sample} by either excluding or including AS714 and
AS753. The error is the $1\sigma$ scatter over $10^5$ random
realizations of the real sample, each generated from a Gaussian
distribution having the above $V_{\rm rms}$ and velocities convolved
with the observational errors.  From the theoretical side, linear
theory for the growth of density fluctuations predicts that the
one--dimensional r.m.s. velocity is 
\be 
V_{\rm
rms}\,=\,{H_0f(\Omega_0)\over \sqrt{3}}\,\left[{1\over
2\pi^2}\int_0^\infty dk\,P(k)\,W^2(kR)\right]^{1/2}\,,
\label{eq:lt}
\ee 
where we use the expression $W(kR)=\exp(-k^2R^2/2)$ for the 
window function.
Croft \& Efstathiou (1994) verified that
eq.(\ref{eq:lt}) provides a rather good fit to results from N--body
simulations for $R$ values in the range 1.5--3$\hm$.
In the present analysis we adopt $R=1.5\hm$, but point
out that the results are largely insensitive to the exact choice. For
instance, assuming $R$ twice as large only increases the final
constraints on $\eta_8$ by about 8\%.

\includegraphics{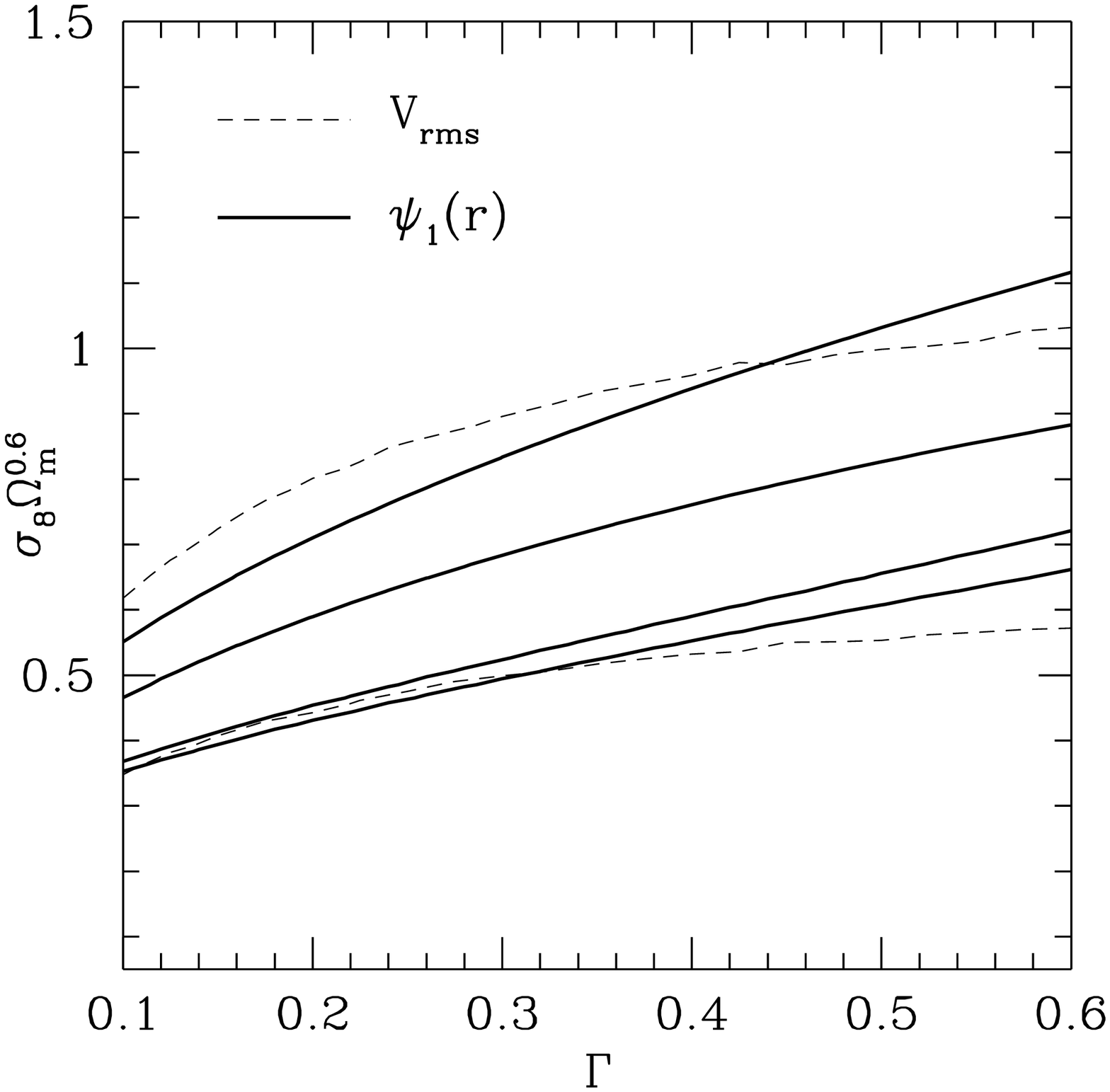}
$~$\\
\vspace{7.7truecm}
$\ \ \ $\\
{\small\parindent=3.5mm {Fig.}~2.--- 
The 1$\sigma$ and $2\sigma$ contours in the
$\eta_8$--$\Gamma$ plane from the analysis of the velocity correlation
function, $\psi_1(r)$, for ENEAR groups and galaxies. Also shown with
the dashed contours is the 1$\sigma$ confidence limits from the
analysis of the r.m.s. peculiar velocity of ENEAR clusters.}

\vspace{5mm}

The procedure to establish the confidence level for a given model is
the same as applied by B97.  Let $v_i$ and $\sigma_i$ be the velocity
and its error for the $i$-th real cluster ($i=1,\dots,16$).  For a
given model, we generate Monte-Carlo samples, each containing 16
velocities, $V_i$, drawn from a Gaussian distribution, with dispersion
provided by eq.(\ref{eq:lt}).  For each sample, every cluster's
velocity is estimated as a Gaussian deviate of the mean $V_{i}$ and
dispersion $\sigma_{i}$, and the r.m.s. velocity of the sample is then
computed.  For each model we generate $N=10^4$ samples and then
compute the fraction ${\cal F}$ with $V^j_{\rm rms}$ ($j=1,\dots,N$)
at least as discrepant as $V^{obs}_{\rm rms}$ with respect to their
average value, $N^{-1}\sum_j V^j_{\rm rms}$.  Therefore, the smaller
the value of ${\cal F}$, the larger the probability for model
rejection. After determining the highest value of ${\cal F}$, relative
confidence levels are computed by determining standard decrements with
respect to this maximum value (i.e., $\Delta {\cal F}\simeq 0.68$ and
0.95 for $1\sigma$ and $2\sigma$ exclusion levels).  The resulting
$1\sigma$ constraints on the $\Gamma$--$\eta_8$ parameter space are
shown in Fig. 2 (dashed curves).  Although this result is less
constraining than that obtained from the velocity correlation
analysis, it nicely overlaps with it, thus demonstrating that ENEAR
clusters and field galaxies consistently trace the same large--scale
flows. For $\Gamma=0.25$ we find $\eta_8=0.63_{-0.19}^{+0.22}$ at the
$1\sigma$ confidence level. This result is also consistent with that
previously obtained from similar analyses of
the SCI cluster velocities (B97, Watkins 1997). The inclusion of the AS714 and AS753
clusters in our analysis would only increase the resulting $\eta_8$ by
about 5\%.

\section{CONCLUSIONS}
\label{conclusions}
We presented statistical analyses of the peculiar velocity field
within $cz=6000\vel$ traced by field objects and clusters in the ENEAR
sample based on $D_n$--$\sigma$ distances. We use the velocity
correlation statistics $\psi_1(r)$ to characterize the velocity field
traced by field ellipticals and loose groups ad find results which are
consistent with those obtained from the SFI sample of spirals with TF
distances. Contrary to past claims, we find no statistically
significant differences between the peculiar velocity fields mapped by
spirals and ellipticals.  This result is in general agreement with and
generalizes the findings of da Costa et al. (2000b) using the
bulk-velocity statistics.
Constraints on the power spectrum of density fluctuations
were derived by resorting to linear theory.
Assuming the
shape of the power spectrum to be consistent with results from galaxy
galaxy clustering analyses, $\Gamma=0.25$, we find
$\eta_8=0.51^{+0.24}_{-0.09}$ at $2\sigma$ level for one interesting
fitting parameter. A consistent constraint is also obtained from the
analysis of the r.m.s. velocity of ENEAR clusters; for
the same value of the shape parameter $\Gamma$, it implies
$\eta_8=0.63_{-0.19}^{+0.22}$ at $1\sigma$, thus consistent with
results from the SCI cluster TF velocities (B97, Watkins 1997).  Our
results confirm the conclusion by B00 that the amplitude of cosmic
flows can be reconciled with independent constraints on the amplitude
of density perturbations as that required by the number density of
nearby rich clusters. They also show that consistent results are
obtainable from independent distance indicators, once they are applied
to homogeneously selected galaxy samples.

\acknowledgments{SB wishes to thank ESO for the hospitality during the
preparation of this work. The authors would also like to thank C. Rit\'e
and O. Chaves for their contribution over the years. The results of
this paper are based on observations at Complejo Astronomico El
Leoncito (CASLEO), Cerro Tololo Interamerican Observatory (CTIO),
European Southern Observatory (ESO), Fred Lawrence Whipple Observatory
(FLWO), Observat\'orio do Pico dos Dias, and the MDM Observatory at
Kitt Peak.}

\end{multicols}


\end{document}